\documentclass[twocolumn,showpacs,preprintnumbers,amsmath,amssymb]{revtex4}

\usepackage{graphicx}
\usepackage{dcolumn}
\usepackage{bm}

\begin{document}

\preprint{APS/123-QED}

\title{Quantum walks on the $N$-cycle subject to decoherence\\ on the coin degree of freedom}

\author{Chaobin Liu}
 \email{cliu@bowiestate.edu}
\author{Nelson Petulante}%
 \email{npetulante@bowiestate.edu}
\affiliation{%
Department of Mathematics, Bowie State University, Bowie, MD, 20715 USA\\
}%
\date{\today}
\begin{abstract}
For a discrete time quantum walk (QW) on the $N$-cycle, allowing for decoherence on the coin, we derive a number of new results, including an explicit formula for the position probability distribution. For a QW of this type, we show that the mixing behavior tends, in the long-run, to a uniform distribution, regardless of the initial state of the system and irrespective of the parity of the number of nodes $N$. These results confirm the findings of previous authors who arrived at similar conclusions through extensive numerical simulations. In particular, we infer that the mixing time $\overline{M(\epsilon)}$ for the time-everaged probability distribution is of order no greater than $O(N^2/\epsilon)$.
\end{abstract}

\maketitle

\section{INTRODUCTION}

When the principles of quantum mechanics get intimate with the theory of Markov Chains, the resulting pedigree is a powerful new paradigm which portends a revolution in the world of electronic computing. Known as the theory of {\em quantum walks} (QW), this novel line of research opens the way to new realms of possibility, including the prospect of super-efficient algorithms capable of treating a class of problems known as ``hard problems" \cite{K03, A03, K06}.  However, even from a purely mathematical perspective, these investigations have inspired important advances in probability theory \cite{K08}. 

Apart from the purely theoretical strides, a number of physical models have been proposed and tested, some of which have managed to induce, with varying degrees of success, QW-like processes in physical systems \cite {DRKB02, KRS03, DLXSWZH03, SBTK03, KRS04, R05, M06, C06, ZDG06, CREG06, P08}. Due primarily to the need to conduct measurements, without which it would not be possible to gather information about the state of the system, all of these proposed models are subject to the ubiquitous phenomenon known as {\em decoherence}. In other words, decoherence is the cost of extracting knowledge about the state of the system. The phenomenon of decoherence has been studied extensively, both numerically and analytically, in various settings, including QWs on a line, on a  cycle, on a hypercube, and graphs of other kinds. For an excellent review, see \cite{K06}. 

In this article, we examine the evolution of a QW on the $N$-cycle under the assumption of decoherence-inducing disturbances on the coin. In general, for a QW on the $N$-cycle, one is interested not only in determining the shape of the limiting (aka {\em stationary}) distribution, but also in estimating the {\em mixing time}. The mixing time refers literally to how fast a stochastic process converges to a stationary distribution. 

QWs on the $N$-cycle first received rigorous treatment in \cite{AAKV01}. Under the idealized assumption of no decoherence-inducing disturbances of any kind and assuming that the parity of the cycle-length $N$ is odd, the authors proved that the coin-governed quantum walk on the $N$-cycle mixes to a uniform distribution. Moreover, they showed that the mixing time for the time-averaged distribution is bounded above by $O(\epsilon^{-3}N\log N)$. Subsequently, these estimates were sharpened in \cite{KT032, R06a, R07}, wherein the mixing time is shown to be of order $O(N\log(1/\epsilon))$. 

More recently, decoherence effects have been imported into the study of QWs on the $N$-cycle. Several models of decoherence have appeared in the literature. Notably, in \cite{BSCR08}, decoherence is modeled by an ambient ``thermal bath" which induces damping of phase and/or amplitude. In this paper, we adopt a model of decoherence which can be described as follows. At each time step of the quantum walk, an observer stands ready to apply a projective measurement. The probability of applying a measurement is given by a fixed parameter $p$, called the ``decoherence rate". In this scenario, three distinct cases merit consideration: decoherence might be assumed to apply to 1) the position only, 2) jointly to the position and coin or 3) to the coin only. In \cite{KT032,R07}, a full analytic treatment is given of case 1) in addition to a numerical treatment of all three cases. In this paper, our main objective is to provide an analytic treatment of case 3).  

In this paper, we adopt an approach employed by Brun {\em et al.} \cite{BCA03, BCA03b} in their treatment of decoherence influences on linear QWs. By adapting this approach to QW's on the $N$-cycle, we derive new insights into of the dynamics of QWs on the $N$-cycle subject to decoherence on the coin degree of freedom. In particular, we provide analytic confirmation of the numerical observations presented by Kendon {\em et al.} in \cite{KT032,K06}. 

\section{Basic Properties of a QW on the $N$-cycle}


For a QW on the $N$-cycle, the {\em position space} of the walker is the Hilbert space $\mathcal{H}_N$ spanned by an orthonormal basis $\{|x\rangle,x \in {\mathbb{Z}_{N}} \}.$  The {\em coin space} is the Hilbert space $\mathcal{H}_2$ spanned by an orthonormal basis $\{|j\rangle, j=-1,1.\}$. The ``state space" is $\mathcal{H} =  \mathcal{H}_N \otimes \mathcal{H}_2$. Thus, a typical state $\psi$ in $\mathcal{H}$ may be expressed as 
$$\psi=\sum_{x \in \mathbb{Z}_{N}}\sum_{j=-1,1}\psi(x,j)|x\rangle\otimes|j\rangle.$$

The evolution of a QW on the $N$-cycle is determined by a unitary operator $U = S(I\otimes C)$, where the ``shift operator" $S : \mathcal{H} \rightarrow \mathcal{H}$ is defined by $S ( |x\rangle\otimes |j\rangle ) = |x+j\rangle \otimes |j\rangle$. As usual, $I$ denotes the identity operator on $\mathcal{H}_{N}$. Meanwhile, any unitary operator $C: \mathcal{H}_2 \rightarrow \mathcal{H}_2$ may serve as the ``coin operator".

Given $\psi_0 \in \mathcal{H}$, let $\psi_t= U^t \psi_0$. Then the sequence of time-iterated
states $\{ \psi_t \}_0 ^\infty$ models the temporal evolution of a pure (totally coherent) QW launched on the $N$-cycle with initial state $\psi_0$. 

Let $X$ denote the position operator on the position space $\mathcal{H}_{N}$, so that 
$X|x\rangle=x|x\rangle$, where $x \in \mathbb{Z}_N$. For a QW with initial state $\psi_{0}$, the probability $P(x,t)$ of finding the walker at the position $x$ at time $t$ is given by the standard formula 
$$P(x,t)=\mathrm{Tr}\left(|x\rangle\langle x|\rho(t)\right),$$
where $\rho(t)=\psi_t \psi_t^{\dagger}$. Thus, at every instant $t$, the eigenvalues of the operator $X_t\doteq {U^{\dagger}}^{t}XU^t$ equate to the possible values of the walker's position with corresponding probability $P(x,t)$.

To conform to reality, any proposed model of quantum computing must be capable of accounting for decoherence-inducing events, including all instances of measurement. Without loss of generality, we are free to assume that these events occur {\em potentially} with probability $p$ at each time step $t$ of the quantum walk. As in [6], the sequence of decoherence-inducing events may be modeled by the probabilistic option of applying to the coin degree of freedom, at each time step of the walk, a {\em unital} family of operators $\{{A}_n\}_{0\leq n\leq \nu}$, jointly satisfying the condition:
\begin{equation} 
\sum_{0\leq n\leq \nu} {\hat{A}}^{\dagger}_n {\hat{A}}_n=I. \label{unital-family}
\end{equation}

Accordingly, when adjusted for decoherence, the ``density operator" of the system acts on the probability density function $\rho$ via the formula:
\begin{eqnarray}
\rho(t+1)=\sum_{0\leq n\leq \nu} U {\hat{A}}_{n} \rho(t) {\hat{A}}^{\dagger}_n U^{\dagger}. \label{decoherence}
\end{eqnarray}

In order to facilitate calculations, including the evaluation of certain fundamental quantities, such as the probability $P(x,t)$, it is advantageous to apply Fourier transformations to all elements of the QW system. The conversion to the Fourier dual amounts simply to a change of basis of the overall state space of the system. 

The conversion begins with the walker's position space $\mathcal{H}_N$, whose ``home basis" of eigenstates is $\{|x\rangle, x = 0, . . . , N - 1\}$. The corresponding Fourier dual is the so-called {\em momentum basis} $\{|k\rangle, k = 0, . . . , N -1\}$, defined explicitly by the formula: 

$$|k\rangle =\frac{1}{\sqrt{N}}\sum_x e^{\frac{2\pi i x k}{N}}|x\rangle.$$

Equivalently, we have:

$$\langle x|k\rangle =\frac{1}{\sqrt{N}}e^{\frac{2\pi i x k}{N}}.$$

Accordingly, the Fourier dual of the evolution operator $U$, denoted by $U_{k}$, is given by 
$$U_k |k\rangle\otimes |j\rangle=C_k |k\rangle \otimes |j\rangle,$$
where \begin{equation}
C_k= \left[\begin{array}{cc}
e^{-\frac{2\pi i k}{N}}& 0 \\
  0   & e^{\frac{2\pi i k}{N}}
\end{array}\right]C.\label{C_k}
\end{equation}

For simplicity, and without loss of generality, we may assume henceforth that every quantum walk under consideration is launched from  position $|0\rangle$ in coin state $|\psi_0\rangle$. As in [6], our analysis utilizes a so-called ``decoherence super-operator" $\mathcal{L}_{kk^{\prime}}$ defined by the formula:
\begin{eqnarray}
\mathcal{L}_{kk^{\prime}}|\psi_0\rangle\langle \psi_0|=\sum_n C_k \hat{A}_n |\psi_0\rangle\langle \psi_0| \hat{A}^{\dagger}_n C_{k^{\prime}}^{\dagger}.\label{superoperator}
\end{eqnarray}

In terms of the super-operator $\mathcal{L}_{kk^{\prime}}$, the formulation of the density operator, as defined in Eq. (\ref{decoherence}) by its action on the density function $\rho$, may be generalized as follows: 
\begin{eqnarray}
\rho(t)=\frac{1}{N}\sum_k\sum_{k^{\prime}}|k\rangle \langle k^{\prime}|\otimes \mathcal{L}^t_{kk^{\prime}}|\psi_0\rangle\langle \psi_0|.
\end{eqnarray}

Similarly, in terms of the super-operator $\mathcal{L}_{kk^{\prime}}$, the formulation of the probability $P(x,t)$ of finding the walker at position $x$ at time $t$ becomes:
\begin{eqnarray}
\!\!\!P(x,t)
\!\!&=&\!\!\mathrm{Tr}\left(|x\rangle\langle x|\rho(t)\right)  \nonumber \\
\!\!&=&\!\!\frac{1}{N}\sum_k\sum_{k^{\prime}}\langle x|k\rangle \langle k^{\prime}|x\rangle \mathrm{Tr}\!\left(\mathcal{L}^t_{kk^{\prime}}|\psi_0\rangle\langle \psi_0|\right)\nonumber \\
\!\!&=&\!\!\frac{1}{N^2}\sum_{k=0}^{N-1}\sum_{k^{\prime}=0}^{N-1}e^{\frac{2\pi i x(k-k^{\prime})}{N}}\mathrm{Tr}\!\left(\mathcal{L}^t_{kk^{\prime}}|\psi_0\rangle\langle \psi_0|\right)\nonumber \\
\!\!&=&\!\!\frac{1}{N}+\frac{1}{N^2}\sum_{k\neq k^{\prime}}e^{\frac{2\pi i x(k-k^{\prime})}{N}}\mathrm{Tr}\!\left(\mathcal{L}^t_{kk^{\prime}}|\psi_0\rangle\langle \psi_0|\right)\!. \label{P(x,t)}
\end{eqnarray}

In the sequel, occasion will arise also to utilize a time-averaged version of ${P(x,\tau)}$. By definition:
\begin{eqnarray}
\overline{P(x,\tau)}=\frac{1}{\tau}\sum_{t=0}^{\tau-1}P(x,t),\label{defofp}
\end{eqnarray}
with the understanding that $P(x,t)$ is given by Eq.(\ref{P(x,t)}). 


To quantify the rate at which the time-averaged distribution $\overline{P(x,\tau)}$ eventually might settle to a stationary distribution, we adopt a measure of transition time $\overline{M(\epsilon)}$ which, for every $\epsilon > 0$, is given by
\begin{eqnarray}
\overline{M(\epsilon)}=\mathrm{min}\left\{\tau\left|\forall t>\tau:\left|\overline{P(x,t)}-P_{\infty}\right|_{\mbox{tv}}<\epsilon\right.\right\}.\label{mixingtime}
\end{eqnarray}
Here $P_{\infty}$ denotes the limiting distribution over the cycle, while the expresion
$$\left|\overline{P(x, t )} - P_{\infty}\right|_{\mbox{tv}} =\sum_x\left|\overline{P(x, t )} - P_{\infty}\right|$$
measures the total variation over the cycle.

Similarly, given $\epsilon > 0$, we define the ``mixing time" $M(\epsilon)$ for $P(x, t)$ by the formula
\begin{eqnarray}
M(\epsilon)=\mathrm{min}\left\{ \tau|\forall t>\tau:\left|{P(x,t)}-P_{\infty}\right|_{\mbox{tv}}<\epsilon\right\} ,\label{mixingtime-1}
\end{eqnarray}
where $P_{\infty}$ and $\|P(x, t ) - P_{\infty}\|_{\mbox{tv}}$ are defined as above.  

Within a margin of error, given by $\epsilon>0$, the mixing time specifies how long it takes for
the time-averaged probability distribution of the walker's position
to transition to its limiting configuration.

\section{Some properties of the decoherence superoperator $\mathcal{L}_{k,k^{\prime}}$ }

To avoid unpleasant complications and to permit us more easily to illustrate our approach to the analysis of a QW on the $N$-cycle subject to decohering influences, we concede, in this paper, to confine our attention to a specific model. Accordingly, to serve as the coin operator of the system, we choose the Hadamard operator: 
\begin{equation}
C_k= \frac{1}{\sqrt{2}}\left[\begin{array}{cc}
e^{-\frac{i2\pi k}{N}}& e^{-\frac{i2\pi k}{N}} \\
  e^{\frac{i2\pi k}{N}}   & -e^{\frac{i2\pi k}{N}}
\end{array}\right].\label{C_k}
\end{equation}

By the same token, to serve as the unital family $\{A_{n}\}_{0\leq n \leq \nu}$ of decoherence-inducing operators on the coin degree of freedom, as in Eq.(\ref{unital-family}), we specialize to the following choice of three ($\nu=2$) operators:

$$\hat{A_0}=\sqrt{1-p}\sigma_0,\,\, \hat{A_1}=\frac{\sqrt{p}}{2}(\sigma_0+\sigma_z),\,\, \hat{A_2}=\frac{\sqrt{p}}{2}(\sigma_0-\sigma_z),$$
where $0\le p \le 1$ and $\sigma_0$ and $\sigma_z$ are the Pauli matrices. The level of decoherence induced by these operators is determined by the value of $p$, called the {\em decoherence rate}. Specifically, the QW evolves as if the state of the coin is measured at each time step with probability $p$. Thus, when $p=0$, the QW evolves as a purely coherent quantum process. At the other extreme, when $p=1$, the QW behaves exactly like a classical random walk.

Now let $\bf{L}(\mathbb{C}^2)$ denote a the Hilbert space of all $2\times2$ complex matrices with inner product given by
\begin{eqnarray}
\langle M_{1},M_{2} \rangle \equiv \mathrm{tr}( M_{1}^{\dagger}M_{2}).
\end{eqnarray}


{Lemma 1.}\,\, Let $\mathcal{S}$ be a superoperator on the Hilbert space $\bf{L}(\mathbb{C}^2)$, defined by 
$$ \mathcal{S}=\sum_{n=0}^2U_1\hat{A}_n\cdot \hat{A}^{\dagger}_nU_2:\,\, B\mapsto \sum_{n=0}^2 U_1\hat{A}_nB\hat{A}^{\dagger}_n U_2,$$
where $U_1$, $U_2$ are $2\times 2$ unitary matrices and $B\in \bf{L}(\mathbb{C}^2)$. Then $\langle \mathcal{S}B, \mathcal{S}B \rangle \leq \langle B,B\rangle$. In particular, $\langle \mathcal{S}B, \mathcal{S}B \rangle = \langle B,B\rangle$ for all $B\in \bf{L}(\mathbb{C}^2)$ if and only if the decoherence rate $p=0$.

Proof. See the Appendix A.

An immediate corollary of this lemma, essential to our analysis, is the fact that $|\lambda|\leq 1$ for every eigenvalue $\lambda$ of $\mathcal{S}$. To justify this, suppose that $B_{\lambda}$ is an eigenvector of $\mathcal{S}$ belonging to $\lambda$. Then $\langle \mathcal{S}B_{\lambda}, \mathcal{S}B_{\lambda}\rangle=\langle \lambda B_{\lambda}, \lambda B_{\lambda} \rangle=|\lambda|^2\langle B_{\lambda}, B_{\lambda}\rangle$. But since, according to the lemma, $\langle \mathcal{S}B, \mathcal{S}B \rangle \leq \langle B,B\rangle$, we see that $|\lambda|\leq 1$.

Now let us specialize to the super-operator $\mathcal{L}_{k, k^{\prime}}$ which maps $\bf{L}(\mathbb{C}^2)$ to $\bf{L}(\mathbb{C}^2)$. If we choose as a basis for $\bf{L}(\mathbb{C}^2)$ the Pauli matrices $\sigma_0$, $\sigma_x$, $\sigma_y$ and $\sigma_z$, then, in terms of this basis, the $4\times 4$ matrix representation of $\mathcal{L}_{k, k^{\prime}}$ is given by: 

\begin{equation}
\mathcal{L}_{k, k^{\prime}}= \left[\begin{array}{cccc}
c^{-}& (1-p)i s^{-}&0 &0 \\
0& 0&(1-p)s^{+} & c^{+}\\
0& 0&(p-1)c^{+} & s^{+}\\
 i s^{-}& (1-p)c^{-}&0 &0
\end{array}\right],\label{L_k}
\end{equation}
where

\begin{eqnarray}
& c^{+}=\cos \frac{2\pi(k^{\prime}+k)}{N},\,\,\, s^{+}=\sin \frac{2\pi(k^{\prime}+k)}{N} \nonumber\\
& c^{-}=\cos \frac{2\pi(k^{\prime}-k)}{N},\,\,\, s^{-}=\sin \frac{2\pi(k^{\prime}-k)}{N}.\nonumber\\
\nonumber\end{eqnarray}
 
After a somewhat tedious, but not very difficult calculation, we arrive at the the following explicit formula for the characteristic polynomial $f(\lambda)$ of $\mathcal{L}_{k, k^{\prime}}$:

\begin{eqnarray}
f(\lambda)&=&\det\left(\lambda \mathbb{I}_{4}-\mathcal{L}_{k, k^{\prime}}\right)\nonumber\\
&=&\lambda^4 
+\left((1-p)c^{+}-c^{-}\right)\lambda^3 
+2(p-1)c^{+}c^{-}\lambda^2 \nonumber \\
&&+(1-p)\left(c^{+}-(1-p)c^{-}\right)\lambda 
+(1-p)^2. \label{eigenpoly}
\end{eqnarray}

The following proposition summarizes some basic attributes of the eigenvalues of $\mathcal{L}_{k, k^{\prime}}$. \\

{Proposition 2.}\,\, Let $\lambda$ be an eigenvalue of $\mathcal{L}_{k, k^{\prime}}$ where $0<p<1$. 

\begin{enumerate}
 \item[(i)] \,$\|\lambda\|\leq 1$;
 
\item[(ii)] \, If $\|\lambda\|=1$ then $\lambda=\pm{1}$; 

\item[(iii)] \, $\lambda=1$ when and only when $k=k^{\prime}$;

\item[(iv)] \, $\lambda=-1$ when and only when $|k^{\prime}-k|=\frac{N}{2}$, in which case the algebraic multiplicity of $\lambda=-1$ is 1.
\end{enumerate} 

 Proof. See Appendix B.
\vskip 0.2in


\section{Evolution of the Hadamard walk on the $N$-cycle subject to decoherence}

As our analysis will show, the behavior of a cyclic QW, when exposed to any level decoherence, however slight, is markedly different from that of a purely coherent QRW. The slightest disturbance forces the QW to behave ultimately like a classical random walk, which, in the long-run, mixes always to a uniform distribution.  Our findings confirm the predictions of \cite{KT032}, which are based on numerical simulations. 



{Theorem 3.}\,\, Suppose a quantum walk, driven by the Hadamard coin operator, is launched on the $N$-cycle with initial coin state $|\psi_0\rangle$ and with decoherence rate $p>0$. If $N$ is odd, then $P(x,t)$ converges to $\frac{1}{N}$ on all nodes of the cycle. If $N$ is even, then $P(x,t)$ converges to $\frac{2}{N}$ on the supporting nodes of the cycle and to 0 on the non-supporting nodes of the cycle.\\
Proof. See the Appendix C.

A classical random walk on a cycle of any size and any parity mixes always to a uniform distribution, both in the strong sense of the raw distribution $P(x,\tau)$ and in the weak sense of the time-averaged distribution $\overline{P(x,\tau)}$. At the other extreme, for a purely coherent QW, the limiting distribution, even in the weak sense of $\lim_{\tau\rightarrow \infty}\overline{P(x,\tau)}$, can fail to be uniform. For instance, consider the special case of a purely coherent QW driven by the Hadamard coin operator. Depending on the parity of the cycle, the limiting distribution, even in the weak sense, may or may not be uniform. According to \cite{AAKV01}, on a cycle with an odd number of nodes, the limiting distribution represented by $\lim_{\tau\rightarrow \infty}\overline{P(x,\tau)}$ is uniform. But, as shown in \cite{TFMK03, BGKLW03}, on a cycle with an even number of nodes, the limiting distribution fails to be uniform unless an extra phase is added to the Hadamard coin operator. 

Our analysis bridges the gulf between the two extremes of purely coherent and purely classical. For a QW on the $N$-cycle driven by the Hadamard coin operator, the following corollary specifies the long-term behavior of the weak limiting distribution represented by $\lim_{\tau\rightarrow \infty}\overline{P(x,\tau)}$.

{Corollary 4.}\,\, Suppose a quantum walk on the $N$-cycle, driven by the Hadamard coin operator, is launched from any initial coin state. If the decoherence rate $p>0$, then the time-averaged distribution $\overline{P(x,\tau)}$ converges to a uniform distribution.

For arbitrary values of $k$, $k^{\prime}$ and $N$, the task of calculating the eigenvalues and eigenvectors of the matrix representation of $\mathcal{L}_{k,k^{\prime}}$ can be quite difficult. This explains our inability, at this time, to offer a more general formula for the mixing time $\overline{M(\epsilon)}$. However, in the special case where $N$ is odd, the following result is relatively easy to justify.     

{Theorem 5.}\,\, Suppose that the quantum walk is released on the $N$-cycle from the initial coin state $\psi_0=|1\rangle$, driven by the Hadamard coin operator and subject to de-coherence rate $p>0$. If $N$ is odd, then the probability $\overline{P(x,t)}$ converges to a uniform distribution with mixing time $\overline{M(\epsilon)}\leq O(N^2/\epsilon)$.

Proof. See Appendix D.

\section{Conclusion and beyond} 

Our analysis suggests a sharp contrast in behavior between quantum walks which are purely coherent and quantum walks which are tainted by even the slightest trace of decoherence. In particular, when exposed to any non-zero level of decoherence on the coin degree of freedom, a QW behaves eventually like a classical random walk. In fact, if the decoherence rate $p>0$, a QW on the $N$-cycle appears to mix always to a uniform distribution at a rate no faster than a classical random walk. However, since our analysis provides only a crude upper bound for the order of the mixing time $\overline{M(\epsilon)}$, the possibility remains open that a significantly sharper upper bound exists. 

Strictly speaking, our results in this paper pertain only to the case of the Hadamard coin operator. However, a similar approach should work at least for the more general type of coin operator $A(\beta)$ treated in \cite{L08}. Moreover, with some further effort, it should be possible to generalize our results by removing any restrictions on the parity of $N$ or the initial state of the coin.

\begin{acknowledgments}

We acknowledge with gratitude the helpful comments offered by the referees. Their suggestions prompted substantial improvements of this paper. Gratefully, we acknowledge the generous support granted to us by the NSF-funded HBCU-UP/BETTER Project at Bowie State University.
 
\end{acknowledgments}

\appendix

\section{PROOF OF LEMMA 1}
Proof. \,\, Suppose $B=(b_{ij})_{2\times2}$. Since $\mathrm{Tr}(M_1M_2)=\mathrm{Tr}(M_2M_1)$, we obtain $\langle \mathcal{S}B, \mathcal{S}B \rangle=(1-p)^2\langle B, B\rangle+(2p-p^2)(|b_{11}|^2+|b_{22}|^2)\leq (1-p)^2\langle B, B\rangle+(2p-p^2)\langle B, B\rangle=\langle B, B \rangle$. The ``='' holds if and only if $(2p-p^2)(|b_{12}|^2+|b_{21}|^2)=0$, which is valid for arbitrary values of $b_{12}$ and $b_{21}$ if and only if $p=0$.

\section{PROOF OF PROPOSITION 2}

Proof of (i).\,\, $\mathcal{L}_{k, k^{\prime}}$ is a special case of the superoperator $\mathcal{S}$ in Lemma 1, according to which, the moduli of all eigenvalues of  $\mathcal{L}_{k, k^{\prime}}$ are less than or equal to unity.

Proof of (ii). \,\, If $e^{i\theta}$ is a non-real eigenvalue of  $\mathcal{L}_{k, k^{\prime}}$, where $\theta$ is a real number, then the conjugate $e^{-i\theta}$ also must be an eigenvalue and $e^{-i\theta}\ne e^{i\theta}$. Hence $f(\lambda)=(\lambda-e^{i\theta})(\lambda-e^{-i\theta})[\lambda^{2}+a\lambda+(1-p)^{2}]$ for some $a\in \mathbb{C}$. Comparing corresponding coefficients of both sides of Eq. (\ref{eigenpoly}), we obtain the following system of equations:
\begin{eqnarray}
a-2\cos \theta&=&(1-p)c^{+}- c^{-}
\nonumber\\
1+(1-p)^2-2a\cos\theta &=& 
-2(1-p)c^{+}c^{-} \nonumber \\ 
a-2(1-p)^2\cos\theta &=&(1-p)c^{+}-
(1-p)^2 c^{-} \label{eqn-1}.\nonumber
\end{eqnarray}
After some elementary algebraic manipulations, we infer that $1+(1-p)^2=-(1-p)\cos\frac{2\pi(k^{\prime}+k)}{N}\cos\frac{2\pi(k^{\prime}-k)}{N}$, which is impossible since the modulus of the LHS is strictly greater than the modulus of the RHS. This contradiction implies that any unit eigenvalue of $\mathcal{L}_{k, k^{\prime}}$ must be real.

Proof of (iii).\,\, $\lambda=1$ is an eigenvalue of $\mathcal{L}_{k, k^{\prime}}$ iff $f(1)=(1-\cos \frac{2\pi(k^{\prime}-k}{N})[1+2(1-p)\cos \frac{2\pi(k^{\prime}+k)}{N}+(1-p)^2]=0$, iff $1-\cos \frac{2\pi(k^{\prime}-k}{N})=0$, which implies $k^{\prime}=k$.

Proof of (iv). \,\, $\lambda=-1$ is an eigenvalue of $\mathcal{L}_{k, k^{\prime}}$ iff $f(-1)=(1+\cos \frac{2\pi(k^{\prime}-k)}{N})[1-2(1-p)\cos \frac{2\pi(k^{\prime}+k)}{N}+(1-p)^2]=0$, iff $1+\cos \frac{2\pi(k^{\prime}-k}{N})=0$, which implies $|k^{\prime}-k|=\frac{N}{2}$. In this case, since $f^{\prime}(-1)=(1-p)^2-1\neq 0$, the algebraic multiplicity of $\lambda=-1$ is 1.

\section{PROOF OF THEOREM 3}

Proof. According to Eq. (\ref{P(x,t)}), 
\begin{equation}
\mbox{}\,P(x,t)\!=\!\frac{1}{N}\!+\!\frac{1}{N^2}\!\sum_{k\neq k^{\prime}}e^{\frac{2\pi i x(k-k^{\prime})}{N}}T_{k k^{\prime}}(t)\label{P(x,t,N)}
\end{equation}
where
\begin{equation}
T_{k k^{\prime}}(t)=\mathrm{Tr}\left(\mathcal{L}^t_{kk^{\prime}}|\psi_0\rangle\langle \psi_0|\right)= 2[1,0,0,0]\mathcal{L}^t_{k,k^{\prime}}\left[\begin{array}{c}
  \alpha_1\\
   \alpha_2 \\
    \alpha_3 \\
     \alpha_4
  \end{array}\right].\nonumber
\end{equation}
We remark that $\alpha_1=\frac{1}{2}$ for all admissible choices of $|\psi_0\rangle$ in the column vector $(\alpha_{1},\alpha_{2},\alpha_{3},\alpha_{4})=[\alpha_{1},\alpha_{2},\alpha_{3},\alpha_{4}]^{T}=|\psi_0\rangle \langle \psi_0|$.

If $N$ is odd and $k\neq k^{\prime}$, then the modulus of every eigenvalue of $\mathcal{L}_{k,k^{\prime}}$ is strictly less than 1, in which case every entry of $\mathcal{L}^t_{k,k^{\prime}}$ tends to zero as $t\rightarrow \infty$. Therefore, $P(x,t)\rightarrow \frac{1}{N}$ as $t\rightarrow \infty$.

It remains to evaluate $P(x,t)$ for $N$ of even parity. For the remaining duration of this proof, let $E_{k k^{\prime}}=e^{\frac{2\pi i x(k-k^{\prime})}{N}}T_{k k^{\prime}}(t)$. Also, for brevity of notation, let both sides of Eq.(\ref{P(x,t,N)}) be multiplied by $N^{2}$. 

By Proposition 2, if $|k-k^{\prime}|=\frac{N}{2}$, then $-1$ is an eigenvalue of $\mathcal{L}_{k,k^{\prime}}$. Accordingly, Eq.(\ref{P(x,t,N)}) becomes:

\begin{eqnarray}
\!\!\!\!\!\!\!\!\!\!\!\!N^{2}P(x,t)
&\!\!=\!\!&{N}+\!\!\!\!\sum_{|k-k^{\prime}|=\frac{N}{2}}\!\!\!\!E_{k k^{\prime}}
 + \!\!\!\!\sum_{|k-k^{\prime}|\neq \frac{N}{2}, \,0}\!\!\!\!E_{k k^{\prime}}\nonumber \\
&\!\!=\!\!&{N}+(-1)^{t}\!\!\!\!\sum_{|k-k^{\prime}|=\frac{N}{2}}\!\!\!\!\cos(\pi x)
+ \!\!\!\!\!\!\sum_{|k-k^{\prime}|\neq \frac{N}{2}, \,0}\!\!\!\!\!\!\!\!E_{k k^{\prime}}.\label{prob}
\end{eqnarray}

The sum of the first two terms of Eq.(\ref{prob}) is either $2N$ or $0$ depending respectively on whether the parities of $t$ and $x$ are equal or opposite. Meanwhile, in the third term of Eq.(\ref{prob}), since $k\neq k^{\prime}$ and $|k-k^{\prime}|\neq \frac{N}{2}$, the modulus of every eigenvalue of $\mathcal{L}_{k,k^{\prime}}$ is strictly less than 1, which implies that every entry of $\mathcal{L}^t_{k,k^{\prime}}$ tends to zero as $t\rightarrow \infty$. Thus, the third term vanishes as $t\rightarrow \infty$. In conclusion, when $N$ is even, $P(x,t)$ tends either to $\frac{2}{N}$ or $0$ as $t\rightarrow\infty$. 

\section{PROOF OF THEOREM 5}

To facilitate the proof of Theorem 5, we invoke a lemma from elementary linear algebra. 

{Lemma.}\,\, Let $A$ be a square matrix with complex entries. If $\lambda=1$ is \underline{not} an eigenvalue of $A$, then, for every positive integer $\tau$, we have $\sum_{t=0}^{\tau-1}A^{t}=(\mathbb{I}-A)^{-1}(\mathbb{I}-A^{\tau})$.

Proof of the Lemma.\, For every positive integer $\tau$, note that $(\mathbb{I}-A)\sum_{t=0}^{\tau-1}A^{t}=\mathbb{I}-A^{\tau}$. Since $1$ is not an eigenvalue of $A$, every eigenvalue of $\mathbb{I}-A$ is non-zero, which implies that $\mathbb{I}-A$ is invertible. Therefore $\sum_{t=0}^{\tau-1}A^{\tau}=(\mathbb{I}-A)^{-1}(\mathbb{I}-A^{\tau})$.

Proof of Theorem 5. Let $[0,\tau]$ denote the time interval over which the quantum walk unfolds. Our strategy is to show, for every value of $x$ in the interval $[0,N]$, where $N$ is the length of the topological cycle, that the value of $\left|\overline{P(x,\tau)}-\frac{1}{N}\right|$ is of order at most $O\left(N/\tau\right)$, where the value of the implicit constant of proportionality is independent of $x$. Although conceptually very simple, the technical aspects of the proof can become quite formidable, unless we exercise due restraint. To minimize the proliferation of excessively lengthy formulas, we introduce the following abbreviations: 

\begin{eqnarray}
\bf{a}&=& 2[1,0,0,0]\nonumber \\
\bf{b}&=& (\mbox{$\frac{1}{2}$},0,0,\mbox{$\frac{1}{2}$})=|\psi_0\rangle \langle \psi_0|\nonumber \\
q &=& 1-p\nonumber
\end{eqnarray}   

We proceed to analyze the expression
\begin{equation}
\overline{P(x,\tau)}-\frac{1}{N}=
\frac{1}{\tau N^2}\sum_{k\neq k^{\prime}}\!\!e^{\frac{2\pi i x(k-k^{\prime})}{N}}\sum_{t=0}^{\tau-1}\!\mathrm{Tr}\left(\mathcal{L}^t_{kk^{\prime}}|\psi_0\rangle\langle \psi_0|\right), \nonumber
\end{equation}
which is based on Eq.(\ref{defofp}). Since $k\neq k^{\prime}$, we know, by Proposition 2, that 1 is not an eigenvalue of $\mathcal{L}_{k,k^{\prime}}$. Therefore, by the preceding Lemma, we have 
\begin{equation}
\sum_{t=0}^{\tau-1}\mathcal{L}_{k,k^{\prime}}^t=(\mathbb{I}-\mathcal{L}_{k,k^{\prime}})^{-1}(\mathbb{I}-\mathcal{L}_{k,k^{\prime}}^{\tau}).\nonumber
\end{equation}
Hence, mindful of the abbreviations introduced above, we have 
\begin{equation}
\sum_{t=0}^{\tau-1}\mathrm{Tr}\left(\mathcal{L}^t_{kk^{\prime}}|\psi_0\rangle\langle \psi_0|\right) 
={\bf a}(\mathbb{I}-\mathcal{L}_{k,k^{\prime}})^{-1}(\mathbb{I}-\mathcal{L}_{k,k^{\prime}}^{\tau}){\bf b},\nonumber 
\end{equation}
in terms of which, the above expression for $\overline{P(x,\tau)}-\frac{1}{N}$ becomes 
\begin{equation}
\frac{1}{\tau N^2}\sum_{k\neq k^{\prime}}e^{\frac{2\pi i x(k-k^{\prime})}{N}}
\cdot {\bf a}(\mathbb{I}-\mathcal{L}_{k,k^{\prime}})^{-1}(\mathbb{I}-\mathcal{L}_{k,k^{\prime}}^{\tau}){\bf b}.\label{mixing}
\end{equation}

Since the modulus of every eigenvalue of $\mathcal{L}_{k,k^{\prime}}$ is strictly less than 1, the $\tau$-th power of $\mathcal{L}_{k,k^{\prime}}$ tends to zero as $\tau\rightarrow\infty$. Thus, for the purpose of estimating the asymptotic behavior of $\overline{P(x,\tau)}-\frac{1}{N}$, we safely may ignore the component of the sum dominated by $\mathcal{L}_{k,k^{\prime}}^{\tau}$, leaving only
\begin{equation}
\frac{1}{\tau N^2}\sum_{k\neq k^{\prime}}e^{\frac{2\pi i x(k-k^{\prime})}{N}}
\cdot {\bf a}(\mathbb{I}-\mathcal{L}_{k,k^{\prime}})^{-1}{\bf b}. \label{d-term}
\end{equation}

After a straightforward calculation, the explicit expansion of ${\bf a}(\mathbb{I}-\mathcal{L}_{k,k^{\prime}})^{-1}{\bf b}$ equates to: 
\begin{equation}
\mbox{}\hspace{-2pt}\sum_{k\neq k^{\prime}}\!\!\frac{1\!-q^2 e^{\frac{i2\pi(k^{\prime}-k)}{N}}\!-q\cos \frac{2\pi(k+k^{\prime})}{N}\!\left[1\!-e^{ \frac{i2\pi(k^{\prime}-k)}{N}}\right]}{\left[1-\cos \frac{2\pi(k^{\prime}-k)}{N}\right]\left[1+2q\cos \frac{2\pi(k+k^{\prime})}{N}+q^2\right]}.\!\!\label{p-1/N}
\end{equation}
The task of finding an upper bound for the modulus of (\ref{p-1/N}) turns out to be quite easy. For each of the summands of (\ref{p-1/N}), note that the modulus of the numerator is bounded above by 4, while the modulus of the second factor (enclosed in square brackets) of the denominator is bounded below by $p^2$. Therefore, the entire sum (\ref{p-1/N}) is bounded above by 
\begin{equation}
\frac{4}{p^2}\sum_{k\neq k^{\prime}} \frac{1}{1-\cos \frac{2\pi(k^{\prime}-k)}{N}}= \frac{8}{p^2}\sum_{j=1}^{N-1}\frac{j}{1-\cos\frac{2\pi j}{N}},\nonumber
\end{equation}
which means that the expression (\ref{d-term}) is bounded above by
\begin{eqnarray}
B(\tau, N)=\frac{8}{p^2\tau N^2}\sum_{j=1}^{N-1}\frac{j}{1-\cos\frac{2\pi j}{N}}. \label{B1}
\end{eqnarray}
Now, provided $N$ is not too small, the value of $B(\tau, N)$ can be estimated by way of a Riemann sum, as follows: 
\begin{eqnarray}
B(\tau,N)&=&\frac{8}{\tau p^2}\sum_{j=1}^{N-1}\frac{\frac{j}{N}.\frac{1}{N}}{1-\cos\frac{2\pi j}{N}}\nonumber\\
&\sim& \frac{8}{\tau p^2}\int_{\frac{1}{N}}^{1-\frac{1}{N}}\frac{udu}{1-\cos (2\pi u)}\nonumber\\
&=&\frac{4}{\tau p^2 \pi^2}\int_{\frac{\pi}{N}}^{\frac{(N-1)\pi}{N}}x\csc^2x dx \nonumber\\
&=& \frac{4}{\tau p^2 \pi^2} (-x\cot x + \ln\sin x){\mbox{\raisebox{-2.5pt}{\huge $|$}}_{\frac{\pi}{N}}^{\frac{(N-1)\pi}{N}}}\nonumber\\
&=&O(N/\tau)\label{orderof-B2}
\end{eqnarray}

Thus, we have 
\begin{eqnarray}
\left|\overline{P(x,\tau)}-\frac{1}{N}\right|\leq O(N/\tau),
\end{eqnarray}
and
\begin{eqnarray}
\left|\overline{P(x, \tau )} - \frac{1}{N}\right|_{\mbox{tv}} 
&=&\sum_x\left|\overline{P(x, \tau )} - \frac{1}{N}\right|\nonumber\\
&\leq& O(N^2/\tau).
\end{eqnarray}
Referring back to Eq. (\ref{mixingtime}), we conclude that the mixing time $\overline{M(\epsilon)}\leq O(N^2/\epsilon)$.

\bibliography{apssamp}

\end{document}